\documentclass[a4paper,11pt]{article}
\usepackage[a4paper,left=3cm,right=3cm,top=3cm,bottom=2.5cm]{geometry}
\usepackage[utf8]{inputenc}
\usepackage{amsmath,amssymb, mathdots}
\usepackage{float}
\usepackage{graphicx}
\usepackage{color}
\usepackage{cite}
\usepackage{hyperref}

\begin{document}

\begin{flushright}
LMU-ASC 28/22\\[2cm]
 \end{flushright}
\begin{center}
{\LARGE\bfseries Towards structures in the flux landscape\\ at large number of moduli}\\[5mm]
\vspace{0.4cm}
Sven Krippendorf${}^{1,2}$, Valent\'i Vall Camell${}^{1}$\\[0.2cm]
{\it ${}^1$ Arnold Sommerfeld Center for Theoretical Physics, Ludwig-Maximilians Universit\"at, Theresienstr.~37, 80333 M\"unchen, Germany\\
${}^2$ Universit\"ats-Sternwarte, Fakult\"at f\"ur Physik, Ludwig-Maximilians Universit\"at, Scheinerstr.~1, 81679 M\"unchen, Germany}
\vspace{1cm}
\end{center}
\begin{abstract}
\noindent 
Sampling string flux vacua enables us to study structures in the string landscape. Here we demonstrate that sampling at large number of moduli is possible for the simplified landscape model of ADK. Using dimensional reduction, we identify analytic structures in these samples. In this example, we find that these structures are rather insensitive to the underlying distribution of UV parameters but they emerge only at large number of moduli and they can be attributed to sampling bias.
\end{abstract}

\newpage

\tableofcontents

\section{Introduction}
Most numerical approaches on studying string vacua focus on examples with low-number of moduli, although the bulk of examples is expected to feature high-dimensional problems. The canonical example being that of studying Calabi-Yau manifolds with small number of moduli. It is currently unknown how do numerical methods scale for these high-dimensional problems\footnote{We thank Andre Lukas for stressing this point at string\_data 2021.}, which we focus on in this paper.

To initiate a systematic analysis for scaling of numerical methods in the string landscape, we focus on the question of identifying a vacuum energy which is within a certain range. This is a prototype of how tunable phenomenological parameters are and how different algorithms introduce a bias on the identified vacua.
The former addresses, for instance, the question on how such landscape of solutions can address the cosmological constant in the standard anthropic sense~\cite{Bousso:2000xa,Susskind:2003kw}. The latter introduces a model for vacuum selection in this landscape which, as we discuss in this work, leads to a selection of different samples. Naturally, this raises the question on how this selection bias compares to biases which are obtained in other phenomenological models of vacuum selection in string theory (e.g.~\cite{Bousso:2007er,Carifio:2017nyb}) and how important they are in comparison to the discrete choice in the underlying landscape which refers to the choice of flux or geometry respectively. Although being an interesting topic, this article focuses only on the methodology of revealing the bias.
We explicitly see that the process of vacuum selection depends not only on the underlying vacuum structure but also on the algorithm. This poses a challenge to any general structures which humans or algorithms do reveal as to understand whether they arise from some selection effect. Given the size of the flux landscape~\cite{Douglas:2003um,Ashok:2003gk,Taylor:2015xtz} and the absence of analytic tools to study universal behaviour of these samples as of today, sampling is necessary to reveal any structures in vacuum solutions of string theory.

In this article, we focus on the setup of Arkani-Hamed, Dimopoulos and Kachru as introduced in~\cite{Arkani-Hamed:2005zuc} which provides a landscape where we can test algorithms in examples with large number of moduli. Although applications on fully-fledged effective field theories arising in flux compactifications on Calabi-Yau orientifolds likely do change the structure in the landscape of vacua, we think that this example provides a non-trivial test for numerical methods. Such a change in structures can also be studied using a random matrix approach~\cite{Marsh:2011aa,Bachlechner:2012at} which we avoid at this moment for simplicity.

The aim of this project is not the construction of efficient algorithms that find vacua with a particular value of the cosmological constant to very high precision as in~\cite{Bao:2017thx} or to study the complexity associated to these algorithms as started in~\cite{Denef:2006ad}, but rather the investigation of how vacua with very close properties are structured inside the space of vacua. This information about structure in the ensemble of vacua can become valuable when trying to develop search algorithms that scale in a tractable manner with the dimension of the space of vacua.

In this landscape we demonstrate that energy based models are also efficient at sampling vacua at large number of moduli, very much independent to the value of the cosmological constant. This confirms the performance of energy based methods previously utilised for examples with small number of moduli (e.g.~\cite{Cole:2019enn,Bena:2021wyr,Krippendorf:2021uxu,Cole:2021nnt} for work on the flux landscape and in a broader string theory context~\cite{Abel:2014xta,Halverson:2019tkf,Larfors:2020ugo,Constantin:2021for,Abel:2021rrj,Loges:2021hvn}).
Further we identify a common structure to these vacua which is mostly insensitive to value of the cosmological constant. Interestingly, this structure is most pronounced at landscapes with a large number of vacua.

The rest of this paper is organised as follows. In Section~\ref{sec:adk} we review the ADK model for the flux landscape and introduce our sampling algorithm. In Section~\ref{sec:structures-in-the-vacua} we present our results on the structures found and provide analytic arguments on why we obtain a polynomial relation among different principal components in our data samples. 
Our conclusions are in Section~\ref{sec:conclusions}.

\section{The ADK model for the flux landscape}
\label{sec:adk}
In the flux landscape -- from a four-dimensional perspective -- we are interested in studying the physics associated to scalar potentials which depend on generically many scalar complex scalar fields and discrete parameters associated to fluxes~\cite{Grana:2005jc,Douglas:2006es}. The distribution of consistent fluxes is constrained, e.g.~from the allowed tadpole, from the UV theory. Although the majority of work has so far focused on the study of type~IIB 
flux compactifications (see for instance~\cite{Martinez-Pedrera:2012teo,Cicoli:2013cha,Demirtas:2019sip} for examples) and the associated closed string moduli, the number of moduli is greatly enhanced when including open string moduli which can be addressed in the language of F-theory. Practically this changes the numerical target from systems of up to a few hundred moduli to systems with a few thousand moduli. As we currently do not have fully-fledged examples, in particular also including open string moduli readily available, it seems useful to test numerical methods on toy examples that can incorporate such scalings such as the model of a vacuum landscape introduced in~\cite{Arkani-Hamed:2005zuc}.

Taking the perspective of random potentials (cf.~\cite{Marsh:2011aa,Bachlechner:2012at}), this example is naturally connected with fully-fledged string constructions as the distribution of random potentials is simply expected to be different. It is these differences which carry some information about what distinguishes potentials arising from a consistent theory of quantum gravity and simple effective field theory potentials.

Given this motivation for the ADK model, we briefly review the relevant properties of this vacuum landscape and introduce our sampling algorithm for this landscape. We also introduce the different choices of parameters used in the rest of the paper.

\subsection{The ADK model}
The model consist of $N$ scalar fields $\phi_i$, each of them with a quartic potential with two minima and with no interaction between them.\footnote{The conventions follow mostly the original literature~\cite{Arkani-Hamed:2005zuc}.} The total potential is then
\begin{equation}
	V = \sum_i^N V_i(\phi_i) \,, \qquad V_i(\phi_i) = \sum_{n=1}^{4} \alpha_{i, (n)} \phi_i{}^n\,.
\end{equation}
Without loss of generality, one can perform field rescalings and shifts on $\phi_i$ such that the potential is of the form
\begin{equation}
	V_i = \alpha_{i, (4)} \phi_i{}^4 + \alpha_{i, (3)} \phi_i{}^3 + \alpha_{i, (2)} \phi_i{}^2 + \alpha_{i, (0)}~.
\end{equation} 
For the terms $\alpha_{i, (0)}$, only the mean of their distribution will contribute to the total potential as a constant. For convenience, we set all of them to $\alpha_{i, (0)}=0$ from now on, though their contribution to $V$ can be easily restored at any point of our discussion. In order to ensure that the potential has two distinct minima for each scalar field we will focus on the region where
\begin{equation}\label{eq:alpha-conditions}
	\alpha_{i, (4)} > 0\,, \qquad \alpha_{i, (3)} \ge 0 \,, \qquad \alpha_{i, (2)} \le 0\,,
\end{equation}
where the choice of sign for $\alpha_{i, (3)}$ is not necessary but can always be taken wlog using the redefinition $\phi_i\rightarrow - \phi_i$. 

For each field $\phi_i$, the two vacua are placed at 
\begin{equation}
	\phi_{i, \pm} = \frac{- 3\, \alpha_{i, (3)} \pm \sqrt{9 \alpha_{i, (3)}{}^2-32\,\alpha_{i, (2)} \alpha_{i, (4)} }}{8\, \alpha_{i, (4)}}\,,
\end{equation}
and we define
\begin{equation}
	V_{i,\pm} = V_i(\phi_{i, \pm})\,.
\end{equation}
The total potential has $2^N$ different vacua corresponding to all distinct choices of vacua for the individual potentials. This in turn implies that $ N\approx 1661$ corresponds to the ubiquitously quoted number of $10^{500}$ flux vacua~\cite{Douglas:2003um,Ashok:2003gk}. Being $\sigma_i(\pm)$ the choice for the field $\phi_i$, the value of the potential at a given vacuum can be written as  
\begin{equation}\label{eq:V-eta-total}
	V_\eta = \sum_{i=1}^N V_{i,\sigma_i(\pm)} = N\,\bar{V}_{{\rm (av)}} + \sum_{i=1}^N \eta_i\, V_{{\rm (diff)},\, i}
\end{equation}
where
\begin{equation}
	V_{{\rm (av)},\, i} = \frac{1}{2}(V_{i,+} + V_{i,-})\,,\qquad 
	V_{{\rm (diff)},\, i} = \frac{1}{2}|V_{i,+} - V_{i,-}|\,,\qquad
	\bar{V}_{{\rm (av)}} = \frac{1}{N} \sum_{i=1}^N V_{{\rm (av)},\, i}\,,
\end{equation}
and $\eta_i$ is a vector in $\mathbb{Z}_2^N$ where each of its entries can take values in $\{+1, -1\}$. Each choice of $\eta_i$ corresponds to a different vacuum of the total potential $V$. For convenience, in our numerical searches we will focus on the second term in~\eqref{eq:V-eta-total} and analyse the quantity
\begin{equation}\label{eq:V-eta-tilde}
	\tilde{V}_\eta = \frac{1}{N} \sum_{i=1}^N \eta_i V_{{\rm (diff)},\,i}\,,
\end{equation}
where we have introduced a  $1 / N$ factor to improve the scaling in this quantity when varying $N.$ Although we will not discuss it further, the quantity $\bar V_{{\rm (av)}}$ can be easily restored. Finally, it is also worth noticing that the maximum value for the quantity $\tilde{V}$ is actually the mean of the $V_{\rm (diff)}$ distribution. Given that we will be using different distributions and we want to compare among them, it will be useful to describe any value of $\tilde{V}$ using a number $\mu \in [-1, + 1]$ as
\begin{equation}\label{eq:mu-description-Vtilde}
	\tilde{V} = \mu\,V_{\rm max}\,, \qquad 
	\text{ with }\quad V_{\rm max} = \langle V_{\rm (diff)} \rangle = \frac{1}{N} \sum_{i=1}^N V_{{\rm (diff)},\,i}\,.
\end{equation}

\subsection{Vacua distributions}
\label{subsec:vacuadistributions}
\begin{figure}
	\begin{center}
		\includegraphics[scale=.45]{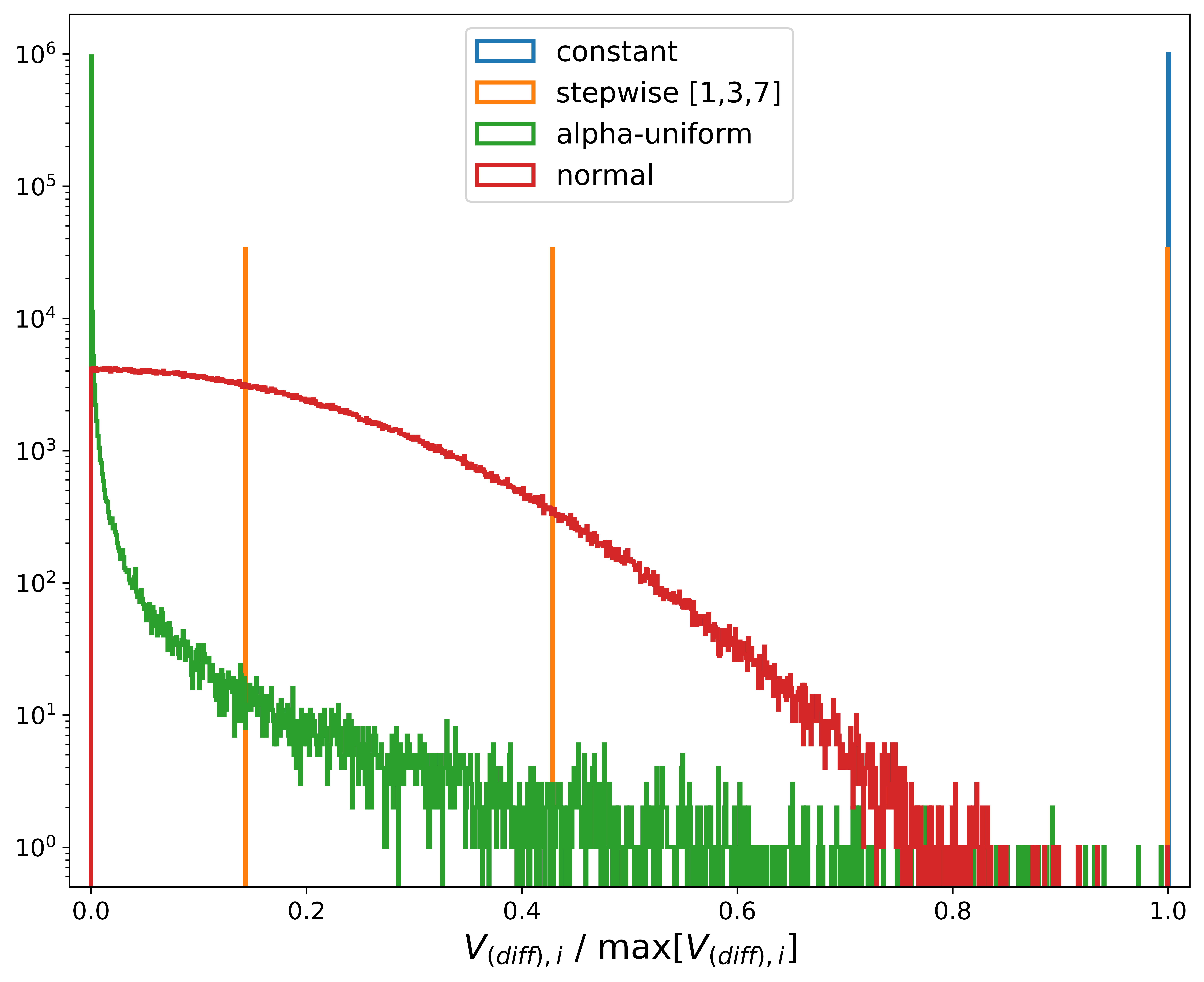}
	\end{center}
	\caption{\label{fig:hist_V_diff_ALL} Histogram of distributions for $V_{{\rm (diff)}, i}$ with $N=10^6$ considered in this work.
The distributions are described in~\ref{subsec:vacuadistributions}. }
\end{figure}

Within the ADK model, the structure of vacua depends on the way $\alpha_{i, (n)}$ -- or equivalently how $V_{{\rm (diff)},\, i}$ --  are distributed. Throughout the paper, we consider different distributions shown in Figure~\ref{fig:hist_V_diff_ALL} which are described subsequently.

\subsubsection*{Constant distribution}\label{sec:vacua-distr-constant}
The simplest distribution possible consist on taking all $\alpha_{i, (n)}$'s with the same value. In particular, we choose them such that
\begin{equation}\label{eq:constant-distribution}
	V_{{\rm (diff)},\, i} = V_0\,,\qquad \forall i\,,
\end{equation}
for a given $V_0$. Although this case appears trivial at first, it will become useful to analyse it and test our sampling algorithms on it, in order to compare the outcomes with those coming from less trivial distributions. 

In this situation, the sets of vacua with the same potential value are trivial to derive. To set some notation, we reproduce the result here. Given the distribution~\eqref{eq:constant-distribution}, the quantity~\eqref{eq:V-eta-tilde} can be written as
\begin{equation}\label{eq:tildeV(v+)-for-constantdist}
	\tilde{V}_\eta = \frac{1}{N}(n_+ - n_-) V_0= \left(\frac{2 n_+}{N} - 1\right) V_0\,,
\end{equation}
where $n_+$ and $n_-$ are the total number of $+1$ and $-1$ in $\eta$ respectively. Therefore, any set $\{\eta\, |\, \tilde V_\eta = \tilde V_{{\rm target}}\}$, for a given $\tilde  V_{{\rm target}}$, is characterized by a number $n_+$, which has to be an integer, and all its elements are related by permutations. 

\subsubsection*{Step-wise constant distribution}\label{sec:vacua-distr-step}
A slightly less trivial case is when the $N$ positions are divided into $P$ classes (which we will consider equally sized for simplicity, assuming also that $N$ is divisible by $P$) each of them with constant value of $\alpha_{i, (n)}$'s but changing from class to class. The distribution of $V_{{\rm (diff)},\, i}$ is then
\begin{equation}
	V_{{\rm (diff)},\, i} = \left\{\begin{array}{cl}
		V_1 &\quad \text{if } i \le \frac{N}{P} \\
		V_2 &\quad \text{if } \frac{N}{p} < i \le \frac{2 N}{P}\\
		\dots & \\
		V_P &\quad \text{if } \frac{(p-1) N}{P} < i \le N
		\end{array} \right. \,,
\end{equation}
with $V_1$, ..., $V_P$ constants.  In this case, 
\begin{equation}\label{eq:step-wise-hypersurface}
\tilde{V}_\eta = \frac{1}{N}\sum_{p=1}^P\left(n^{(p)}_+ - n^{(p)}_-\right)V_p = 
\frac{2}{N}\sum_{p=1}^P n^{(p)}_+V_p - \frac{1}{P}\sum_{p=1}^P V_p\,,
\end{equation}
where $n^{(p)}_+$ is the number of $+1$ in class $p$. For a fixed $\tilde{V}_\eta$, equation~\eqref{eq:step-wise-hypersurface} defines a hypersurface in a $P$-dimensional space. Therefore, the set $\{\eta\, |\, \tilde V_\eta = \tilde V_{{\rm target}}\}$ is described by all points $(n_+^{(1)},\dots n_+^{(P)})$ which lay in the corresponding hypersurface and satisfy the conditions $n_+^{(p)}\in \mathbb{Z}$ and $0\ge n_+^{(p)}\ge N/P$ for all $p=1,\dots, P$. All $\eta$ vectors with the same $(n_+^{(1)},\dots n_+^{(P)})$ are related by permutations inside each class.

\subsubsection*{Uniform distribution on $\alpha_{i, (n)}$}\label{sec:vacua-distr-uniform}
A more general case is to have all $\alpha_{i, (n)}$'s to be different. We consider a situation were the $\alpha_{i, (n)}$ are sampled from a uniform distribution. Following~\eqref{eq:alpha-conditions} we take
\begin{equation}\label{eq:uniform-alpha-distribution}
	\alpha_{i, (2)} \sim \mathcal{U}(-1, 0)\,,\qquad 
	\alpha_{i, (3)} \sim \mathcal{U}(0,1)\,,\qquad
	\alpha_{i, (4)} \sim \mathcal{U}(\alpha_{\text{cutoff}}, 1)\,,
\end{equation}
where $\alpha_{\text{cutoff}}$ is a cutoff to avoid values of $\alpha_{i, (4)}$ being too close to zero, in which case positions with such small values would dominate over the others. For $N\ge 10^4$, it is sufficient to have $\alpha_{\text{cutoff}} \sim 10^{-2}$ in order that any set of $n$ potential-values $\{V_\eta\}$ obtained by randomly choosing $n$  $\eta$-vectors and computing the corresponding potential approximates a normal distribution for $n$ large enough. For $N=10^6$, the resulting distribution of $V_{{\rm (diff)},\,i}$ obtained after sampling $\alpha_{i, (n)}$ according to~\eqref{eq:uniform-alpha-distribution} is plotted in Figure~\ref{fig:hist_V_diff_ALL}.

\subsubsection*{Normal distribution on $V_{{\rm (diff)},\, i}$}\label{sec:vacua-distr-normal}
Finally, we will also consider the case where we take $V_{i,\pm}$ from normal distributions, namely
\begin{equation}
	V_{i,\pm} \sim \mathcal{N}(0,1)\,.
\end{equation} 
Note that, in this case, the distribution of the positive values $V_{\rm (diff)}$ will be a half Gaussian, as shown in Figure~\ref{fig:hist_V_diff_ALL}. We observe that, compared to the uniform distribution of $\alpha_{i, (n)}$, 
it has a much smaller peak close to zero but also a slower decay. This distribution is used in~\cite{Arkani-Hamed:2005zuc, Bao:2017thx}.

\subsection{The sampling algorithm}
In the following sections, we will study the properties of sets of configurations with similar value of the total potential. To construct these sets, we use a slight adaptation of the Metropolis algorithm, which we review in Appendix~\ref{sec:Metropolis_algorithm}.

Given a distribution of $V_{{\rm (diff)},\,i}$, the aim of the algorithm is to find $\eta$ vectors, $\eta\in\mathbb{Z}_2^N$, with a value of $\tilde{V}_\eta$ in $\tilde{V}_{{\rm target}} \pm \delta\tilde{V}_{{\rm target}}$. The sampling algorithm will be characterised by the following distribution and parameters:
\begin{equation}\label{eq:algorithm-parameters}
	V_{{\rm (diff)}, i}\,,\quad 
	N\,,\quad
	n_{{\rm samples}}\,, \quad 
	\tilde{V}_{{\rm target}} \,, \quad
	\delta\tilde{V}_{{\rm target}}\,, \quad
	n_{{\rm reshuffle}}\,, \quad
	n_{\rm max.\, steps}\,, \quad
	\sigma_{{\rm Metropolis}}\,,
\end{equation}
where $N$ is the number of fields, or equivalently the dimension of vectors $\eta$. A key aspect of our algorithm is that, once a sample is found, the algorithm does not restart from a completely far away point (which would in general make it slow) but only $n_{\rm reshuffle}$ randomly chosen positions are flipped before the searching algorithm starts again. As we will discuss in the rest of the paper, this parameter will play an important role on how the resulting samples are distributed. Finally, $n_{\rm max.\, steps}$ is the number of steps after which the algorithm will stop if it is unable to find any sample and $\sigma_{{\rm Metropolis}}$ is the standard deviation of the normal distribution  in the Metropolis step.  Further details are described in Appendix~\ref{sec:Metropolis_algorithm}.
\begin{table}
 	\begin{center}
 	\renewcommand{\arraystretch}{1.5}
	\begin{tabular}{|l|l|}
		\hline
		Parameter & values \\
		\hline\hline
		$N$ & $10^3$ - $10^6$\\
		\hline
		$n_{{\rm samples}}$ & $100$ - $200$ \\
		\hline 
		$\tilde{V}_{{\rm target}}$ &  $\mu = 0$ - $0.85$
		\\
		\hline
		$\delta\tilde{V}_{{\rm target}}$ & $10^{-9}$ - $10^{-10}$ \\
		\hline
		$n_{{\rm reshuffle}}$ & $10 ^ 2$ - $N$ \\
		\hline 
		$n_{\rm max.\, steps}$ & $10 ^ 6$ \\
		\hline
		$\sigma_{{\rm Metropolis}}$ & $100 \cdot \delta\tilde{V}_{{\rm target}}$\\
		\hline
	\end{tabular}
    \renewcommand{\arraystretch}{1}
	\end{center}
\caption{\label{tab:parameters-summary} Summary of values of the parameters~\eqref{eq:algorithm-parameters} used in the analysis of Section~\ref{sec:structures-in-the-vacua}.}
\end{table}

In the following sections we will use this sampling algorithm to construct sets of $\eta$-samples with similar $\tilde{V}_\eta$ for different distributions of $V_{{\rm (diff)},\,i}$ and analyse the structure of these vacua using principal component analysis. The values of the algorithm parameters that we scan through are summarised in Table~\ref{tab:parameters-summary}.

Note that sampling of our vacua is rather efficient while already using this Metropolis algorithm. More efficient methods could become relevant when identifying nearby vacua becomes less efficient.

\section{Structure in the vacua}\label{sec:structures-in-the-vacua}
As sampling turns out rather efficient in our search for all distributions previously introduced, we directly move to analysing the structure in the vacua. 
We show that all of them have a limit at large $N$ where samples organise following some universal structures, and argue subsequently how these are related to permutations.

\begin{figure}
	\includegraphics[scale=.195]{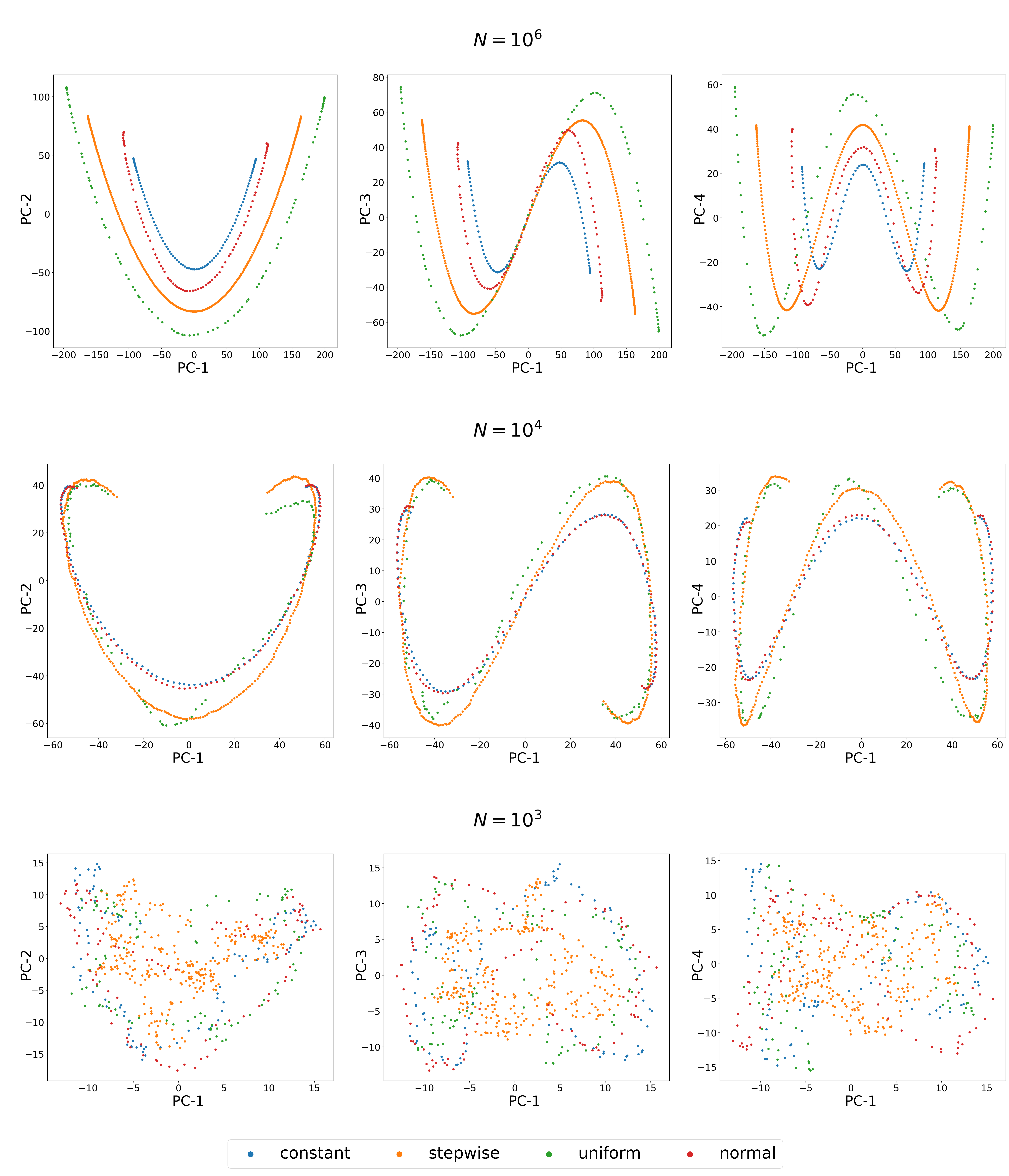}
	\caption{\label{fig:PCA-all-dist} Polynomial relationship among principal components (PCs) in our vacua samples ({\bf left:} first and second PC, {\bf middle:} first and third PC, {\bf right:} first and fourth PC). Different curves show the results for different distributions introduced in Section~\ref{subsec:vacuadistributions}. Principal component analysis sets of samples obtained using the Metropolis algorithm  on $V_{{\rm (diff)},\,i}$ distributions of section~\ref{subsec:vacuadistributions} with different values of $N$ and all with fixed $n_{{\rm reshuffle}} = 100$. The target value is $\mu=0.1$ in the notation of~\eqref{eq:mu-description-Vtilde}, with $\delta\tilde{V} = 10^{-10}$ in the continuous cases, and the PCA is performed independently on each set.}
\end{figure}

On a sample of vacua, we perform principal component analysis which reveals uniformly across our different distributions patterns summarised in Figure~\ref{fig:PCA-all-dist}. We see that at large $N,$ polynomial dependences between the first and subsequent PCA components are present (top row of Figure~\ref{fig:PCA-all-dist}). This structure is slightly deformed for intermediate $N$ which is shown in the second row and it disappears for small $N$. All these diagrams are at a fixed value of $n_{\rm reshuffle}.$ We find that larger $n_{\rm reshuffle}$ values also render the polynomial relationships between the PCA components to disappear. Naturally this raises the question  {\it which significance shall we attribute to this structure?} Foremost this structure enables the generation of distinct vacua satisfying our phenomenological requirement of a vacuum energy within a pre-defined range. This leaves the opportunity to utilise such structure in sampling algorithms directly, which at this stage we do not need to explore, but, depending on the complexity of sampling, might become relevant. For moderate $n_{\rm reshuffle}$ values compared to $N$, our sampling algorithm is utilising, as we argue below, a permutation symmetry among different fields. Coarsely speaking, the algorithm is able to compensate for changes in one of the vacuum energies with a switch of the vacuum energy for another field. At large values for $n_{\rm reshuffle}$ compared to $N$, this structure due to the permutation symmetry is firstly deformed and then with increasing ration of $n_{\rm reshuffle}$ over $N$ it disappears.

To discuss this behaviour in detail, we start with the case of the constant distribution of vacuum energies and subsequently discuss the other distributions. Details on additional checks can be found in Appendix~\ref{app:extra-data}.

\subsection{Constant distribution}
As described in section~\ref{sec:vacua-distr-constant}, the sets of samples with fixed $\tilde{V}_{{\rm target}}$ can be constructed analytically. However, it is useful to apply our search algorithm in this space to have a better understanding of how it performs and compare it with less trivial cases. 

If one uses the algorithm with a big ratio between the number of fields $N$ and the number of positions that are randomly changed after finding a sample, $n_{\rm reshuffle}$, the set of samples seems to organize along a curve in $\mathbb{R}^N$ (rigorously speaking,  after embedding the set of points in $\mathbb{Z}_2^N$ into $\mathbb{R}^N$). For example, the results of plotting the first four principal components against the first one for a sample obtained with $N=10^6$ and $n_{{\rm reshuffle}}=100$ can be found in the first line of plots of Figure~\ref{fig:PCA-all-dist}. In particular, we observe polynomial-like shapes with the degree of the polynomial increasing with the component of the PCA analysis. 

The reason why this structure appears can be understood analytically. Suppose we have the following set with $n$ samples for a target value $\tilde{V}_{{\rm target}} = N^{-1}(n_+ - n_-):$
\begin{equation}\label{eq:circulant-set}
	\begin{split}
	\Bigl\{ & \Bigl(\overbrace{1,\dots, 1 }^{n_+}, \overbrace{-1,\dots, -1 }^{n_-}), \\
	& \Bigl(\overbrace{-1,... -1}^{m}, \overbrace{1,..., 1 }^{n_+}, \overbrace{-1,..., -1 }^{n_- - m}\Bigr), \\
	& \Bigl(\overbrace{-1,... -1}^{2 m}, \overbrace{1,\dots, 1 }^{n_+},\overbrace{-1,..., -1 }^{n_- - 2m}\Bigr), \\
	& \dots \\
	& \Bigl(\overbrace{-1,... -1}^{(n-1)\, m}, \overbrace{1,\dots, 1 }^{n_+},\overbrace{-1,..., -1 }^{n_- - (n-1)\, m}\Bigr)\Bigr\}\,,
	\end{split}
\end{equation}
where $m$ is an integer satisfying $(n-1)\, m \le n_-$. For the sake of argument we will also assume that $n_+ \ge n_-$ (all arguments can be adapted to the situation $n_+ < n_-$ by transforming $\eta\rightarrow - \eta$). This implies that $(n-1)\, m \le n_+$, which means that the first $n_+$ positions, which are all $(+1)$'s in the first sample, can  become all $(-1)$'s only in the last sample of the set and only in the case when $n_+=n_-$ and a multiple of $m$. 

Seen as a matrix, each row in~\eqref{eq:circulant-set} is constructed by shifting the previous one $m$ positions to the right. Such a symmetry allows for an analytic principal component analysis, as we discuss in Appendix~\ref{app:SVD-circulant-set}. There we show that the result of plugging the $n$-th principal component of such set against its first one is a polynomial of degree $n$, which appear because we are plotting two cosine functions of different frequencies against each other. 

The fact that PCA on the vacua samples we obtain using Metropolis algorithm resembles the same polynomials is because the  resulting set is actually very similar to~\eqref{eq:circulant-set} after positions are reorganized. To get a better intuition, let us analyse the structure of~\eqref{eq:circulant-set} a bit more closely. Between the first and the second sample of the set~\eqref{eq:circulant-set}, $m$ $(+1)$'s and $m$ $(-1)$'s are flipped. This positions are left untouched for the rest of the set. Then, from second to third another $m$ $(+1)$'s and $m$ $(-1)$'s are flipped and remain like this for the rest of the set, and analogously for the rest of the samples. A possible way to visualise this is by looking at the mean of the value of each position through the set of samples. Discarding the positions which do not change  along the set of samples (which do not play any role while performing PCA), it is easy to see that the means along the different columns of the set~\eqref{eq:circulant-set} will be uniformly distributed within the following values
\begin{equation}\label{eq:means-distribution}
	\frac{1}{n}(-n + 2)\,,\quad \frac{1}{n}(-n + 4)\,,\quad \dots \quad \frac{1}{n}(n - 2)\,.
\end{equation}

Let us now compare with the set of samples obtained from Metropolis algorithm. We consider a situation where $n_{{\rm samples}},\, n_{{\rm reshuffle}} \ll n_-,\, n_+$ and assume we have selected a target $\tilde{V}_{{\rm target}}$ for which equation~\eqref{eq:tildeV(v+)-for-constantdist} has a solution for an integer $n_+$. Suppose we are in a situation where the algorithm has already found its first sample. Then, the next step is to randomly flip $n_{{\rm reshuffle}}$ positions to generates a new $\eta$ away from $\tilde{V}_{{\rm target}}$. In the worst case where all flipped positions are different, along this process we will have flipped approximately $n_{\rm reshuffle}\times n_+ \times N^{-1}$ positions which had a value of (+1)  and  $n_{\rm reshuffle}\times n_- \times N^{-1}$ positions with a value of (-1) and the value of $\tilde{V}$ will have  moved to approximately
\begin{equation}
	\tilde{V}_{{\rm target}} \rightarrow \tilde{V}_{{\rm target}} - \frac{2\,n_{\rm reshuffle}}{N}\,\tilde{V}_{{\rm target}} \,.
\end{equation}
At this point the search part of the algorithm starts again. This implies flipping other (or maybe some of the same) positions in order to find another sample for $\tilde{V}_{{\rm target}}$. Note that, each time we flip a position, the value of $\tilde{V}$ changes by $\Delta\tilde{V} = \pm 2 N^{-1} V_0$. Therefore, the minimal number of flips that the algorithm will have to perform to have again a sample with value $\tilde{V}_{{\rm target}}$ is $|\mu_{{\rm target}}|\times n_{{\rm reshuffle}}$, where $\mu_{{\rm target}}$ describes our target value of $\tilde{V}$ as in~\eqref{eq:mu-description-Vtilde}. This is in fact a very good estimation when we take our algorithm with $\sigma_{\rm Metropolis}\rightarrow 0$, in which case the probability of keeping a flip that moves us away from the target is almost negligible.

With all these considerations we conclude that the probability of a position that has changed from the first to the second sample to change again in the third can be estimated to be  $\lesssim (1 + |\mu_{\rm target}|)~N^{-1}\times \, n_{{\rm reshuffle}}$. Considering that this procedure is done $n_{{\rm samples}}$ times, the probability for each of the first flipped positions to remain unchanged along the full set of samples is then $\gtrsim \bigl[1 - (1 + |\mu_{\rm target}|)~N^{-1}\times \, n_{{\rm reshuffle}}\bigr]^{n_{{\rm samples}}}$.  For the case in the first line of Figure~\ref{fig:PCA-all-dist}, this probability is $98.9\%$.

\begin{figure}
	\includegraphics[scale=.3]{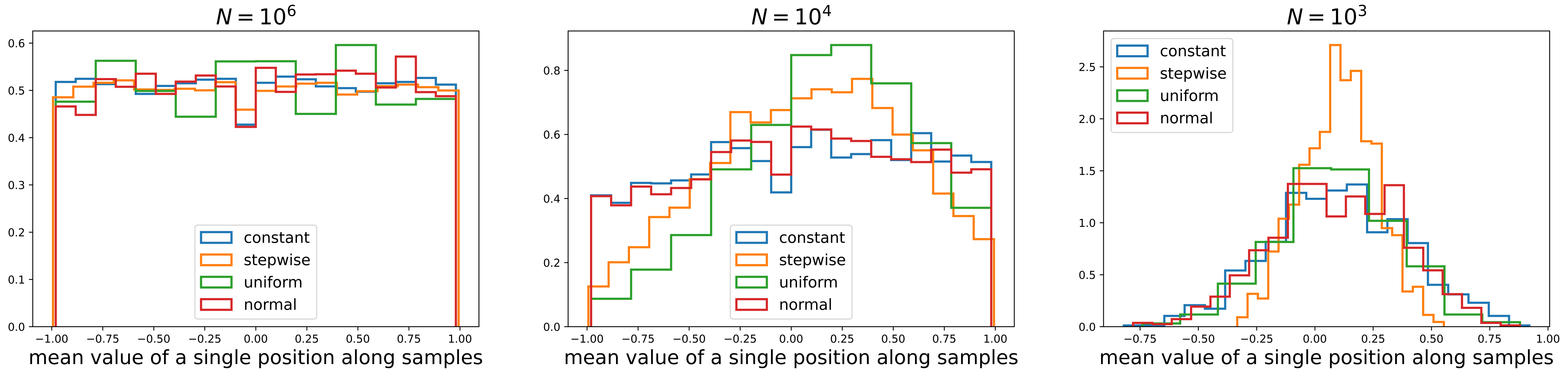}
	\caption{\label{fig:hist-means-all-dist} Empirical confirmation that the mean value of individual vacuum energy contributions remains uniform for large $N$ across different distributions ({\bf Left:} $N=10^6$). 
 All these sets have a target $\mu=0.1$ in the notation of~\eqref{eq:mu-description-Vtilde} and $\delta\tilde{V} = 10 ^{-10}$ in the continuous cases. They all have fixed $n_{{\rm reshuffle}} = 100.$ In the plots we ignore the positions with mean  $+1$ or $-1$. The {\bf middle} ($N=10^4$) and {\bf right} ($N=10^3$) plots show the change in the distributions.}
\end{figure}

One way to empirically check that the set resulting from our Metropolis algorithm is similar to~\eqref{eq:circulant-set} is by computing the mean of the value of each position along the samples and plotting a histogram. The result can be found in the first plot of Figure~\ref{fig:hist-means-all-dist}. We observe that the resulting distribution is close to a uniform one, as expected from our previous arguments.

Finally, one can explore how our algorithm behaves away from the limit where the ratio between $N$ and $n_{{\rm reshuffle}}$ is very large. Looking at the last two rows of Figure~\ref{fig:hist-means-all-dist}, we observe that, as we move away from this limit, the structure we have defined so far gets deformed towards a cloudy distribution of the data. At the same time, the distribution of the mean values of the different positions moves away from the uniform distribution towards a normal one, as can be seen in the plots of Figure~\ref{fig:hist-means-all-dist} in the middle and on the right, indicating that the obtained set is not similar to~\eqref{eq:circulant-set}. It is worth noticing that, for the case with  $N=10^4$ and $n_{{\rm reshuffle}} = 100$, despite we are clearly away from the situation described by~\eqref{eq:circulant-set}, the PCA shows that our samples still organise along some curves, though these are clearly not polynomial. We leave the study of this situation for future work.

\subsection{Step-wise constant distribution}
In the previous section we have tested our sampling algorithm in a trivial setup and identified a limit at large ratio between $N$ and $n_{{\rm reshuffle}}$ where the algorithm produces sets analogous to~\eqref{eq:circulant-set}. We now argue that this limit also exists in situations where the distribution of $V_{\rm (diff)}$ is much less trivial. 

\begin{figure}
	\begin{center}
		\includegraphics[scale=.36]{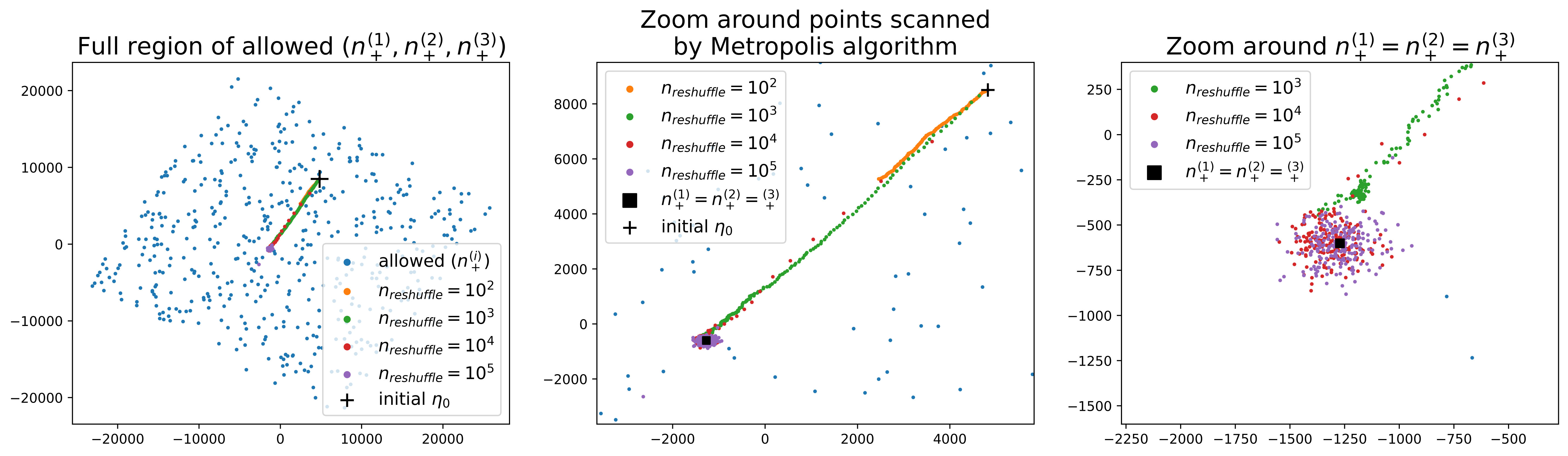}
	\end{center}
	\caption{\label{fig:n+plane_stepwisedist} Projection (using PCA) of points laying on the 2-plane of allowed  $(n_+^{(1)}, n_+^{(2)}, n_+^{(3)})$ for target with $\mu=0.1$, $N=10^5$ and $V_{{\rm (diff)},\, i}$ given by~\eqref{eq:step-dist-137}. The blue points of the  first plot correspond to 500 different randomly chosen allowed $(n_+^{(1)}, n_+^{(2)}, n_+^{(3)})$ combinations. The dots in other colours are those obtained when using  Metropolis algorithm with different values of $n_{{\rm reshuffle}}$. All different runs of our algorithm start from the same randomly chosen initial $\eta_0$ (marked as a black cross), which for convenience is taken to have $\tilde{V}=\tilde{V}_{target}$ . {\bf Left:} Full  $(n_+^{(1)}, n_+^{(2)}, n_+^{(3)})$ plane. {\bf Center:}~Zoom around the points obtained by our search algorithm. {\bf Right:} Zoom around the point $n_+^{(1)} =  n_+^{(2)} = n_+^{(3)}$, marked as a black squared in the last two plots.}
\end{figure}

As a first generalisation, we consider a step-wise constant distribution described in Section~\ref{sec:vacua-distr-step}. In this case we still have analytical control over the solutions but not all permutations of the positions lead to configurations with the same $\tilde{V}_{{\rm target}}$. For the concrete experiments of this section, we will consider the following distribution:
\begin{equation}\label{eq:step-dist-137}
V_{{\rm (diff)},\, i} = \left\{\begin{array}{cl}
1 &\quad \text{for } i \le \frac{N}{3} \\
3 &\quad \text{for } \frac{N}{3} < i \le \frac{2 N}{3}\\
7 &\quad \text{for } \frac{2 N}{3} < i \le N
\end{array} \right. \,.
\end{equation}
As discussed in section~\ref{sec:vacua-distr-step}, a proper way to characterise configurations in this situation is by the set of numbers $(n_+^{(1)}, n_+^{(2)}, n_+^{(3)})$, corresponding to the number of $(+1)$'s the configuration has in its class. For configurations corresponding to the same $\tilde{V}_{{\rm target}}$, these points are located in a two-dimensional plane within $\mathbb{R}^3$, which is depicted as the blue dots in Figure~\ref{fig:n+plane_stepwisedist} for the target $\mu=0.1$, following the notation of~\eqref{eq:mu-description-Vtilde}.

We next use our sampling algorithm in the vacua distribution~\eqref{eq:step-dist-137}. We start by considering a limit with large $N$ and small $n_{{\rm reshuffle}}$, concretely\footnote{As discussed in Section~\ref{sec:vacua-distr-step}, we take all classes to be equal size. Therefore we effectively take $N=10^6 - 1$. The same observation holds for all other searches in this section.} $N=10^6$ and  $n_{{\rm reshuffle}}=100$. The result of the PCA on the obtained set is shown in the first row of Figure~\ref{fig:PCA-all-dist}. Again we observe the appearance of polynomial shapes similar to those obtained in the constant distribution case of the previous section.

Let us then try to understand to which extend the set we obtained resembles~\eqref{eq:circulant-set}. As a first empirical check, we again compute how the means of the value of each position across samples are distributed. The results for the case $N=10^6$  and $n_{{\rm reshuffle}}=100$ can be found in the left plot of Figure~\ref{fig:hist-means-all-dist}. Again, we observe that, for the limit where the ratio between $N$ and $n_{{\rm reshuffle}}$ is big, the distribution gets close to the uniform distribution of~\eqref{eq:circulant-set}.

Next, we exploit further the fact that we still have analytical control over the sets of configurations with the same $\tilde{V}_{{\rm target}}$ to gain better intuition about how our algorithm behaves. In particular, we are interested in knowing which portion of the set of all possible configurations with the same $\tilde{V}_{{\rm target}}$ our algorithm is able to explore. To this aim, it is useful to keep track of which classes $(n_+^{(1)}, n_+^{(2)}, n_+^{(3)})$ the algorithm visits. The results are shown in Figure~\ref{fig:n+plane_stepwisedist}, where we show the sets obtained by our algorithm starting from the same randomly chosen initial $\eta_0$ and choosing different values of $n_{{\rm reshuffle}}$. We observe that, for all choices of $n_{{\rm reshuffle}}$, the algorithm tends to move towards the region close to the point where $n_+^{(1)} = n_+^{(2)} = n_+^{(3)}$. For small $n_{\rm reshuffle}$, the algorithm moves very slowly and it is not able to reach the mentioned point with a small number of samples. When $n_{\rm reshuffle}$ is big, it moves faster and can reach this region, from which it is not able to move away any more. Therefore, even in the case where $n_{\rm reshuffle}$ is very large, the algorithm explores only relatively small regions of the space of all possible solutions.

Let us now analyse why this happens. Suppose we start from an initial $\eta_0$ with random $n_+^{(i)}$. For the sake of the argument, we assume that $\tilde{V}(\eta_0) < \tilde{V}_{{\rm target}}$. Next we start the search part of the algorithm which, given that it chooses positions at random, will visit each class approximately the same amount of time. However, given that we need to increase $\tilde{V}$, the algorithm will have a greater impact on those classes with smaller $n_+^{(i)}$, increasing their value, until a sample is found. Then, the algorithm shifts $n_{{\rm reshuffle}}$ positions, chosen randomly and therefore affecting equally each class. Similar to the case of the previous section, the numbers $n_+^{(i)}$ change approximately as 
\begin{equation}
	n_{+}^{(i)} \rightarrow  \left(1 - \frac{2\, n_{\rm reshuffle}}{N}\right)n_{+}^{(i)} + \frac{n_{{\rm reshuffle}}}{3}
\end{equation}
and the change $\Delta n_+^{(i)} = n_+^{(i)}{}_{(new)} - n_+^{(i)}{}_{(old)}$ is therefore
\begin{equation}
	\Delta n_+^{(i)} \sim - \frac{2\,n_{{\rm reshuffle}}}{N}\left(n_+^{(i)} - \frac{N}{6}\right)
\end{equation}
which is negative if $n_+^{(i)} > N/6$ and positive otherwise. The absolute value of the change is proportional to $n_{{\rm reshuffle}}$, as one could have expected, and also on how away from the middle point $n_+^{(i)}$ is, therefore having a greater effect on extreme $n_+^{(i)}$ values. Combining both parts of the algorithm we  can then conclude that in general it tends to have greater impact on those classes with extreme values of $n_+^{(i)}$ until a situation is reached where $n_+^{(1)} \sim n_+^{(2)} \sim n_+^{(3)}$. In this situation, the algorithm affects all classes equally on average, an therefore the algorithm cannot escape this region. This is compatible with the result we observe in Figure~\ref{fig:n+plane_stepwisedist}: all samples try to move to the point of equal $n_+^{(i)}$, those with bigger $n_{{\rm reshuffle}}$ can move faster due to the bigger changes during the reshuffling. Once the algorithm reaches a region around the point of equal  $n_+^{(i)}$, it cannot leave it. 

Finally, let us comment on what happen when we move away from the limit of big ratio between $N$ and $n_{{\rm reshuffle}}$. As  discussed, even in the situation with big $n_{{\rm reshuffle}}$, the algorithm cannot scape the region around the point of equal $n_+^{(i)}$ and explore the space of all possible configurations with the same $\tilde{V}$. In this situation, however, given that the ratio between $N$ and $n_{{\rm reshuffle}}$ becomes smaller, the samples stop being organised as in~\eqref{eq:circulant-set}, as one observes from the mean analysis in the last two plots of  Figure~\ref{fig:hist-means-all-dist}. Performing PCA on these sets of samples, we observe that this corresponds to the deformation and eventually disappearance of the polynomial structures, as can be seen in the last two rows of Figure~\ref{fig:PCA-all-dist}.

\subsection{Uniform distribution}
We next explore how our search algorithm behave in a non-trivial distribution of $V_{\rm (diff)}$. In this case, the values $V_{{\rm (diff)}, i}$ are randomly chosen from the distribution described in section~\ref{sec:vacua-distr-uniform}. Given that the values of the individual $V_{{\rm (diff)}, i}$  are real and the distribution has compact support due to the cut-off $\alpha_{\text{cutoff}}$, there will always be one sample within $\tilde{V}_{{\rm target}}\pm \delta\tilde{V}$ for any $\tilde{V}\in[-\tilde{V}_{\rm max}, \tilde{V}_{\rm max}]$ and $\delta\tilde{V}$ if one takes $N$ to be large enough. Here, however, we will not focus on finding single examples with very small $\delta\tilde{V}$, but instead observe how our algorithm is able to find samples with similar $\tilde{V}$ and how they organise. 

We start again by applying our search algorithm on the distribution of section~\ref{sec:vacua-distr-uniform} with a big ratio between $N$ and $n_{{\rm reshuffle}}$. In particular, we begin by taking $N=10^6$ and $n_{{\rm reshuffle}}=100$ and consider samples with $\tilde{V}_{{\rm target}} = 0.1\, V_{\rm max} \pm 10^{-10}$. The result of the PCA on the obtained set can be again found in the first row of Figure~\ref{fig:PCA-all-dist}. Again, we observe the appearance of shapes that resemble the  polynomial structures encountered in the previous sections, although we are now in a situation where permutations are not a symmetry.

To have a better understanding of why these shapes appear, we analyse how similar the set of obtained samples is to~\eqref{eq:circulant-set}. To this aim, we check again the the distribution of the means of the different positions along samples. The results of this analysis can be found in Figure~\ref{fig:hist-means-all-dist}. We observe that, similar to what happened in the previous cases, in the situation where the ratio between $N$ and $n_{{\rm reshuffle}}$ is big the distribution resembles a uniform one, getting deformed into a Gaussian when one moves away from this limit.

\begin{figure}
	\begin{center}
		\includegraphics[scale=.45]{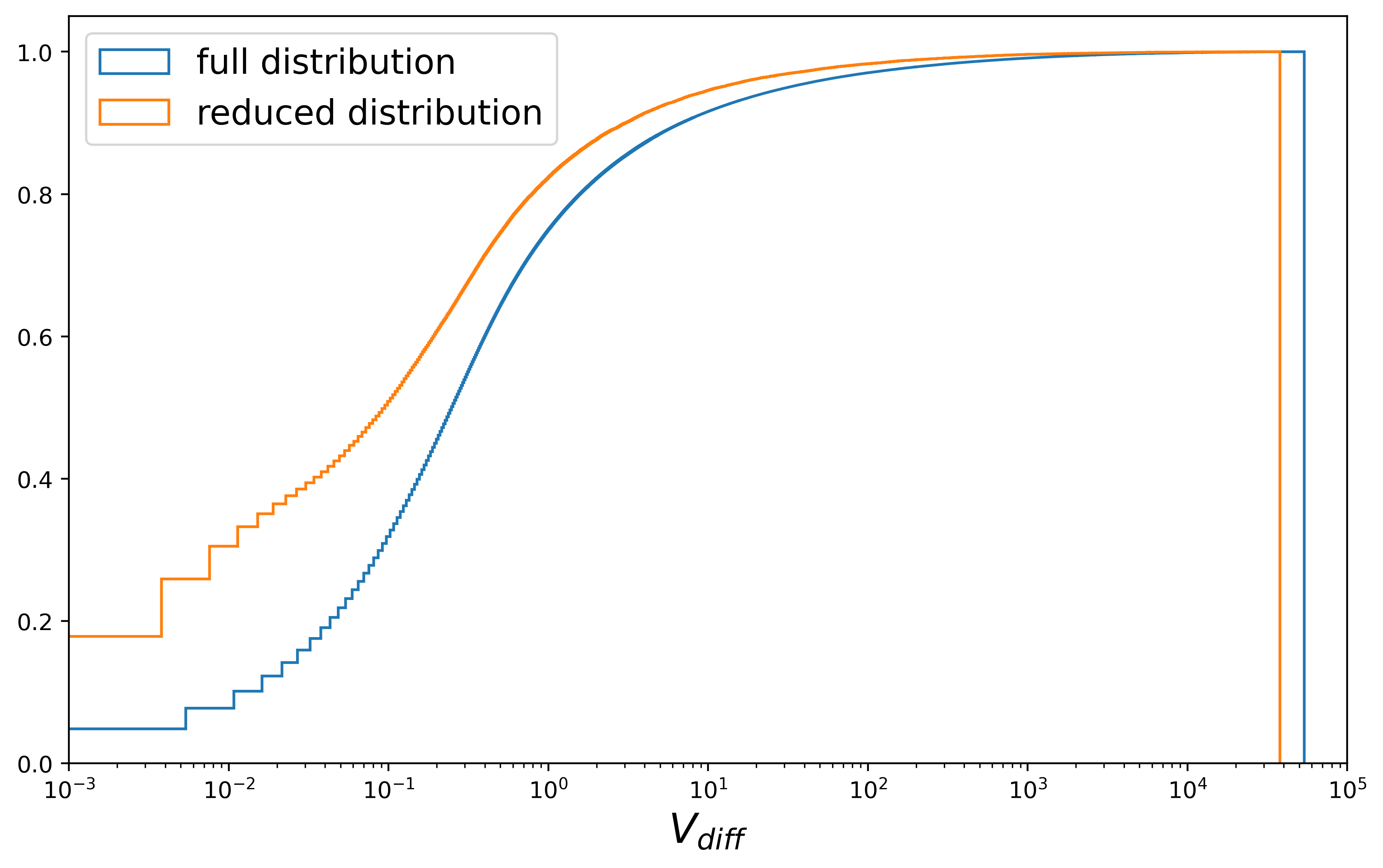}
	\end{center}
	\caption{\label{fig:density_hist_Vdiff_10^6_reduced} Density of cumulative histogram of different distributions $V_{{\rm (diff)},\,i}$ to show the relative importance of components with smaller or higher vacuum energy contributions. The curve indicates what percentage of the total distribution is below a certain value of $V_{\rm (diff)}$. The blue line is obtained as described in section~\ref{sec:vacua-distr-uniform} with $\alpha_{\text{cutoff}}=10^{-2}$ and $N=10^6$, while the orange line corresponds to the same distribution after removing the values of the positions that remain unaffected during our sampling. 
	}
\end{figure}

From the analysis above follows that, even in the case when the distribution of $V_{{\rm (diff)}, i}$ is far from constant, in a certain limit we still obtain sets that are similar to~\eqref{eq:circulant-set}. To gain more intuition why this is happening, we next analyse what are the values $V_{{\rm (diff)}, i}$ of the positions that have been changed during our search algorithm. We do this by taking the distribution of $V_{{\rm (diff)}, i}$ and removing the values corresponding to the positions that remain unchanged during the whole performance of the sampling algorithm, and compare the resulting distribution with the original one using cumulative density histograms. These indicate which percentage of the distribution lays below a given value of  $V_{\rm (diff)}$. The results are shown in Figure~\ref{fig:density_hist_Vdiff_10^6_reduced}. We observe that the plot for the reduced distribution is always above the original one, indicating that our sampling algorithm tends to affect positions with smaller value of $V_{\rm (diff)}$. 

In fact, if one focuses on the region of small $V_{\rm (diff)}$, the situation starts resembling those analysed in the previous two sections. In particular, looking at the  distribution obtained by sampling the $\alpha_i$ parameters from a uniform distribution plotted in green in Figure~\ref{fig:hist_V_diff_ALL}, one observes that this is a very populated region. Actually, in this region it is possible to find positions whose $V_{{\rm (diff)},i}$ value differ by less than $10^{-4}$. Note that, when computing $\tilde{V}$ as in~\eqref{eq:V-eta-tilde}, we divide the contributions of  $V_{{\rm (diff)},i}$ by $N$. Therefore, taking $N=10^6$, the contributions to the total $\tilde{V}$ of two position that differ by less than $10^{-4}$ will differ by less than $10^{-10}$, below our error threshold.  
Therefore, it is reasonable to think  that the polynomial shapes that we observe in the first line of Figure~\ref{fig:PCA-all-dist} for the uniform distribution of the $\alpha$ parameters have a strong relation with permutations along these sets of values of $V_{\rm (diff)}$.

Finally, we also explore how our algorithm behaves as we go away from the limit where the rate between $N$ and $n_{{\rm reshuffle}}$ is large. As we move away, the shapes start to get deformed and eventually disappear. At the same time, the resulting sets start looking less similar to~\eqref{eq:circulant-set}, until being eventually very different. The results for changing $N$ while keeping the same $n_{{\rm reshuffle}}$ can be seen in the last two lines of ~\ref{fig:PCA-all-dist} and the last two plots of~\ref{fig:hist-means-all-dist}. In this case, microscopically, when $N$ is reduced the distribution is less populated, reducing the number of positions with very closed values of $V_{\rm (diff)}$. The structures also get deformed if one keeps a big $N$ but increases the value of  $n_{{\rm reshuffle}}$, as one can see in Figure~\ref{fig:PCA-uniform-N=10^6-different-mu} in appendix~\ref{app:extra-data}.

\subsection{Normal distribution}
Finally, we apply our algorithm to another non-trivial distribution of $V_{\rm (diff)}$, in this case a normal distribution as described in section~\ref{sec:vacua-distr-normal}. 

We analyse again the case where $N$ is much bigger than $n_{{\rm reshuffle}}$. In particular we take $N=10^6$ and $n_{{\rm reshuffle}} = 100$. We look for values $\tilde{V} = 0.1 V_{\rm max} \pm 10^{-10}$. Performing PCA on the set of obtained samples,  we observe again the polynomial structures, as shown in the first line of Figure~\ref{fig:PCA-all-dist}.

Performing similar analysis as in the previous section we observe that, even though the distributions are different, our search algorithm performs in a very similar way in this limit, showing a universal behaviour. In particular, the means of the value of the positions along samples are again approximately uniformly distributed in the limit of big ratio between $N$ and $n_{{\rm reshuffle}}$  and get deformed to a normal distribution when we move away from this limit, as can be seen in Figure~\ref{fig:hist-means-all-dist}. Although the normal distribution from which we sample $V_{{\rm (diff)}\,i}$ in this section is very different from the other distributions used so far, it is still possible to find regions which are very populated, making it possible to use permutations to give an explanation for the appearance of the polynomial shapes in the limit of big ratio between $N$ and $n_{{\rm reshuffle}}$.

\section{Conclusions}
\label{sec:conclusions}
As numerical explorations of string theory solutions are gaining momentum, several methodological challenges arise with it. Here we touch upon two of these challenges:
\begin{enumerate}
\item How do our numerical methods scale with the respective dimension of the system?
\item How can we reliably obtain structures from our samples?
\end{enumerate}
To address the first challenge, we investigate the large $N$ limit of the ADK model and look for configurations with a certain value of the total potential in situations where the solutions are not sparse. We use a simple Metropolis algorithm to look for samples. Once a solution is found, instead of starting from a completely random point, we change only $n_{{\rm reshuffle}}$ randomly chosen positions. Since we are starting from a closer point, the algorithm finds the next sample faster.

To identify structures in our samples, we study how the resulting samples are distributed using PCA. We observe three types of behaviour depending on how big $n_{{\rm reshuffle}}$ compared to $N$. When $n_{{\rm reshuffle}}$ is small compared to $N$, we observe the appearance of polynomial-like shapes. These get deformed if one increases the value of $n_{{\rm reshuffle}}$ until getting a cloud distribution when $n_{{\rm reshuffle}}$ is close to $N$. We argue that the polynomial-like shapes can be related to certain permutations of positions that have very similar individual potentials. We find that these shapes also disappear when we look at independent samples, i.e.~we increase the ratio $n_{\rm reshuffle}/N.$ We argue that these behaviours are universal, i.e.~independent of the particular distribution for the individual potentials, as far as this has compact support or decrease rapidly, allowing for regions that become very populated when we increase $N$. 

It remains an open question whether this information can be used to design algorithms that have a better scaling with $N$? For example, by constructing an algorithm that takes into account that certain permutations transform samples in the desired range of value for their total potential into new samples within the same desired range.

Ending on a positive note, we seem to have reached the era where structures in string theory datasets can be revealed using data science approaches and it is intriguing what this direction will reveal when analysing large datasets constructed in the past decades of string theory research.

\subsection*{Acknowledgements}
The authors thank Andreas Schachner and Yanick Thurn for useful discussions. The work of VVC is supported by the Alexander von Humboldt Foundation via a Feodor Lynen fellowship. We acknowledge the Mainz Institute for Theoretical Physics (MITP) of the Cluster of Excellence PRISMA+ (Project ID 39083149) for hospitality and support during part of this work.

\appendix

\section{Metropolis algorithm} \label{sec:Metropolis_algorithm}
In this appendix we review the Metropolis algorithm used to obtain the samples analysed in section~\ref{sec:structures-in-the-vacua}. Using the parameters~\eqref{eq:algorithm-parameters}, the algorithm performs as follows
\begin{enumerate}
	\item Start with an initial vector $\eta = \eta_0$ and compute $\tilde{V}_\eta$ using~\eqref{eq:V-eta-tilde}. Call this quantity $\tilde{V}_{\eta_{old}}$.
	
	\item Randomly choose one position in $\eta$ and flip its sign. Compute the potential associated with the new $\eta$, $\tilde{V}_{\eta_{new}}$, using~\eqref{eq:V-eta-tilde}. Then do the following:
	\begin{enumerate}
		\item If $\tilde{V}_{\eta_{new}} \in \tilde{V}_{{\rm target}} \pm \delta\tilde{V}_{{\rm target}}$:  jump to step 4.
		\item If $|\tilde{V}_{\eta_{new}} - \tilde{V}_{{\rm target}}| <  |\tilde{V}_{\eta_{old}} - \tilde{V}_{{\rm target}}|$ keep the flip of sign and redefine $\tilde{V}_{\eta_{new}}$ as $\tilde{V}_{\eta_{old}}$. 
		\item If the above condition is not satisfied, then keep the flip only with probability $P$, given by
		\begin{equation}
		P = \frac{p(\tilde V_{new})}{p(\tilde V_{old})}\,,\qquad \text{ where } 
		p(\tilde{V}_\eta)\sim e^{-\frac{1}{2}\left(\frac{\tilde{V}_\eta - \tilde{V}_{{\rm target}}}{\sigma_{Metr.}}\right) ^ 2}\,.
		\end{equation} 
		If the flip is kept, redefine $\tilde{V}_{\eta_{new}}$ as $\tilde{V}_{\eta_{old}}$. Otherwise undo the flip and go back to the previous $\eta$. 
	\end{enumerate}
	
	\item Iterate step 2. $n_{\rm max.\, steps}$ times or until condition (a) is hit. 
	
	\item If condition (a) is hit, check if the corresponding $\eta$ vector is stored in your samples set. If not, store it. Then, randomly choose $n_{{\rm reshuffle}}$ positions in $\eta$ and flip their signs. Call this vector the new $\eta_0$ and redo steps 1. - 3.
	
	\item Stop the algorithm either when your samples set contains $n_{{\rm samples}}$ distinct elements or whenever step 2. is iterated $n_{\rm max.\, steps}$ times in a row without hitting condition (a).
\end{enumerate}

\section{Principal component analysis for translation-generated data}\label{app:SVD-circulant-set}
\subsection{Principal component analysis and singular value decomposition}
Consider a data matrix $X$ containing one sample point per row. $X$ has dimensions $(n,\,N)$, where $n$ is the number of samples and $N$ is the dimension of our feature space, and for simplicity we assume $n<N$, which is the situation in the main text, although our arguments are easily generalisable.  Let us also assume that the data is centred, namely each feature (column) has mean zero across all samples. 

The principal component analysis (PCA) of a set $X$ consists on transforming it into an orthogonal basis where the first basis vector is along the direction of higher variance and the $i$-th vector is along the direction that maximises the variance while being orthogonal to the first $(i - 1)$ ones.  If the data set $X$ is centred, this is equivalent to
\begin{equation}\label{eq:pca-def}
	T = X\cdot W\,,
\end{equation}
where $T$ is the transformed data matrix and  $W$ is a matrix whose columns are the eigenvectors of the square matrix $(X^T\cdot X)$ ordered in decreasing order of the corresponding eigenvalue. Note that $(X^T\cdot X)$ is semi-positive definite and therefore such order always exist. 

The singular value decomposition (SVD) of a rectangle matrix $X$ is given by
\begin{equation}\label{eq:svd-def}
	X = U\cdot \Sigma\cdot W^T\,,
\end{equation}
where $U$ and $W$ are squared matrices of appropriate dimensions and $\Sigma$ a $n\times N$ matrix with $\Sigma_{ii}\ge 0$ for $i=0,\dots n$ and zero otherwise. The matrices $U$ and $W$ are in fact the eigenvector matrices of $(X\cdot X^T)$ and $(X^T\cdot X)$  respectively and the non-zero values in $\Sigma$, called singular-values, are the square root of their eigenvalues. If $X$ is centred, the matrix $W$ is the same appearing in~\eqref{eq:pca-def}. From now on, we will assume that $U$, $W$ and $\Sigma$ are ordered decreasingly according to the corresponding singular-values.

Finally, combining~\eqref{eq:pca-def} and~\eqref{eq:svd-def} one obtains
\begin{equation}
	T = U\cdot \Sigma\,,
\end{equation}
and we conclude that the values the data along the $i$-th principal component are given in the $i$-th eigenvector of $(X\cdot X^T)$ multiplied by the $i$-th singular value.

\subsection{SVD for translation-generated data}
In this section we will solve the eigenvalue problem for the matrix $(X\cdot X^T)$ in the case where $X$ is generated as in~\eqref{eq:circulant-set}. We start discussing the case where $m=1$. After removing the columns that do not change along the samples (the ones on the right in~\eqref{eq:circulant-set} and some in the middle if $n\,m \le n_+$) one obtains the following data matrix
\begin{equation}
	\tilde X = \left(\begin{array}{cccccccccc}
	1 & 1 &1 & \dots& 1 &-1 & -1& -1& \dots & -1 \\
	- 1 & 1 & 1 & \dots & 1 & 1 &-1&-1& \dots & -1 \\
	- 1 & -1 & 1 & \dots & 1 & 1 &1&-1& \dots & -1 \\
	\vdots \\
	- 1 & -1 & -1 & \dots & -1 & 1 &1&1& \dots & 1 \\
	\end{array}\right)
\end{equation}
which is a matrix of dimension $n \times 2\,(n-1)$, where $n$ is the original number of samples in~\eqref{eq:circulant-set}, and each line contains exactly $(n-1)$ $(+1)$'s and $(n-1)$ $(-1)$'s. This can be seen as a translation-generated matrix in the sense that $\tilde X_{i+1, j} = \tilde X_{i, (j+1 \text{ mod } 2(n-1))} $ .

In order to compute perform PCA, one needs to centre the data matrix. The mean along the columns are
\begin{equation}
	m = \frac{1}{n}\Bigl(- (n-2), -(n-4), \dots, (n - 4), (n - 2), (n - 2), (n - 4), \dots, -(n-4), -(n - 2)\Bigr)\,,
\end{equation}
consistent with the distribution discussed in~\eqref{eq:means-distribution}. The centred data matrix is them
\begin{equation}
	X = \tilde{X} - \mathbf{1}\otimes m \,,
\end{equation}
where $\mathbf{1}$ is a $n$-dimensional vector filled with $1$'s. It is straightforward to check that
\begin{equation}
	J_{(n)}\cdot X \cdot J_{(2(n-1))} = X
\end{equation}
with $J_{(k)}$ a $k$-dimensional square matrix with ones along the anti-diagonal and zero otherwise,
\begin{equation}
	J_{(k)} = \left(\begin{array}{ccc}
	0 & \cdots & 1 \\
	 \vdots&\iddots &\vdots\\
	 1 &\cdots & 0
	\end{array}\right)\,.
\end{equation}
It then follows that $J_{(n)}$ commutes with $X\cdot X^T$, implying that both can be diagonalised simultaneously. Given that $J_{(n)}$ has eigenvalues $\pm 1$, all eigenvectors $u$ of $X\cdot X^T$ will satisfy $J\cdot u = \pm u$.

The eigenvectors are
\begin{equation}
	u_{(k)} = \left\{\begin{split}
	\big(u_{(k)}\big)_p & = \cos\left( \frac{-1+2\,p}{2n}\,k\,\pi \right)
	\quad \text{ for } i=1,\dots (n-1)\\
	\big(u_{(k)}\big)_p & =1\quad \forall p \qquad \text{ for } i=n
	\end{split}\right.
\end{equation}
where $p = 1,\dots, n$ are the components of the vector.  The corresponding eigenvalues are
\begin{equation}
	\lambda_{(k)} =\frac{4}{3 n}   \sec \left(\frac{\pi  k}{2 n}\right) \sum _{j=1}^{n} \left[\left(1 + 3 (j-1) j + 3 (1-2 j) n+2 n^2\right) \cos \left(\frac{ -1 + 2 j}{2 n}\,k\,\pi\right)\right].
\end{equation}
Using the recursive formula
\begin{equation}
	\cos\big((n+2)\, x\big) = 2 \cos\big((n+1)\,x\big) \cos x - \cos(n x)\,,
\end{equation}
it is easy to see that the $p$-th component of the vector $u_{(k)}$ can always be written as a polynomial of degree $k$ of the $p$-th component of the vector $u_{(1)}$. Combining this with the discussion in the previous section provides an analytical explanation for the appearance of polynomial shapes when performing PCA on the set~\eqref{eq:circulant-set} with $m=1$. This result can be easily generalised to the case with arbitrary $m$ by realising that
\begin{equation}
	X^{(m)}\cdot (X^{(m)}) ^T = m\, X^{(1)}\cdot (X^{(1)}) ^T\,,
\end{equation}
where $X^{(m)}$ denotes the case of arbitrary $m$ and $X^{(1)}$ the case with $m=1$. Therefore, the matrix $U$ for the case of arbitrary $m$ is the same as in the case with $m=1$ while the principal values get rescaled by $\sqrt{m}$.

\section{Additional checks}\label{app:extra-data}
\begin{figure}[H]
	\includegraphics[scale=.35]{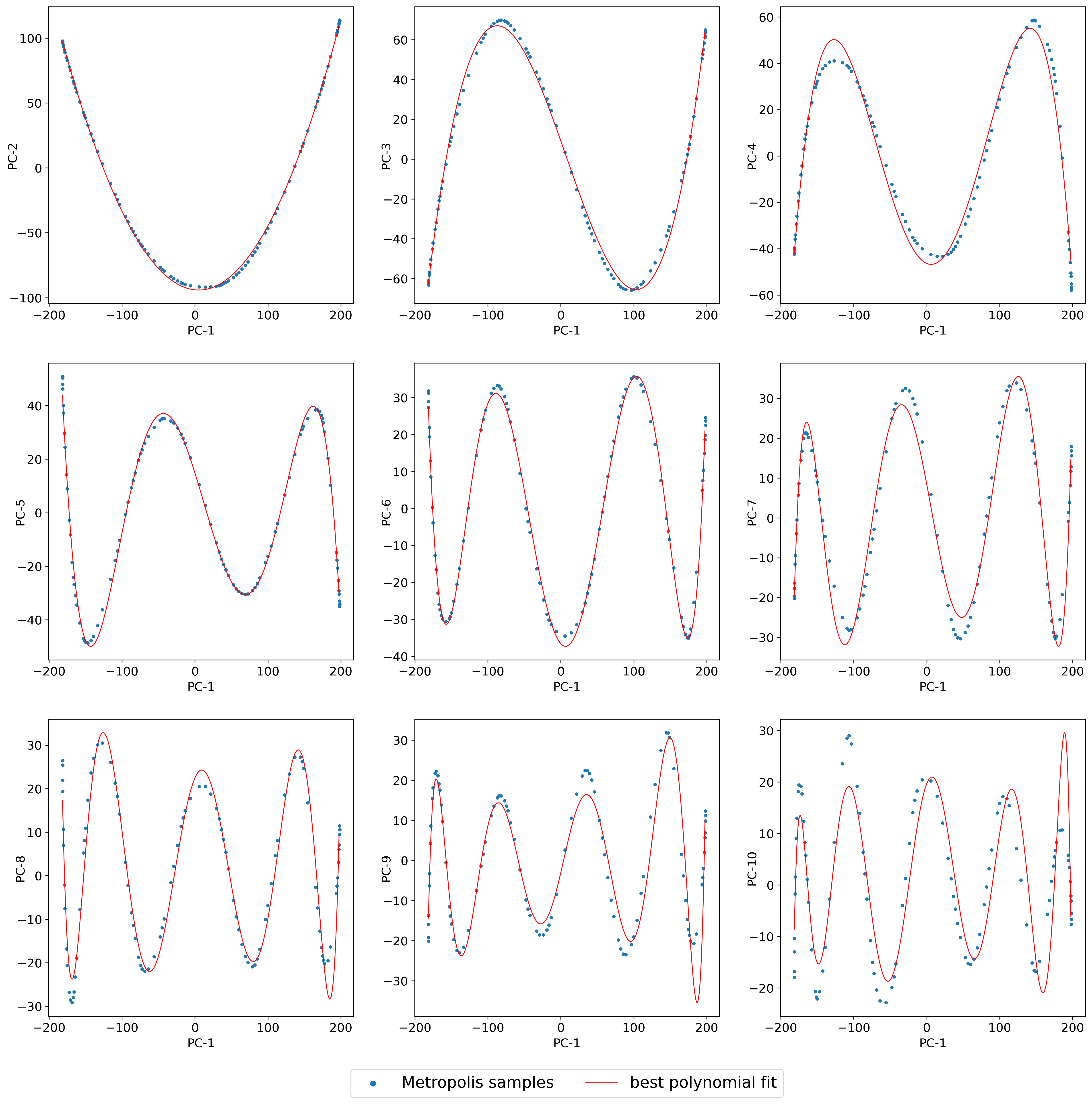}
	\caption{\label{fig:PCA-uniform-N=10^6-n=100} Principal component analysis of a set of samples obtained using the Metropolis algorithm on the uniform distribution of section~\ref{sec:vacua-distr-uniform} with $N=10^6$, $\tilde{V}_{{\rm target}} = 0.1$, $\delta\tilde{V}_{{\rm target}} = 10^{-10}$, $n_{{\rm reshuffle}} = 100$. Red lines are the best polynomial fits with degrees being increased from 2 (top-left) to 10 (bottom-right).}
\end{figure}

\begin{figure}[H]
	\includegraphics[scale=.195]{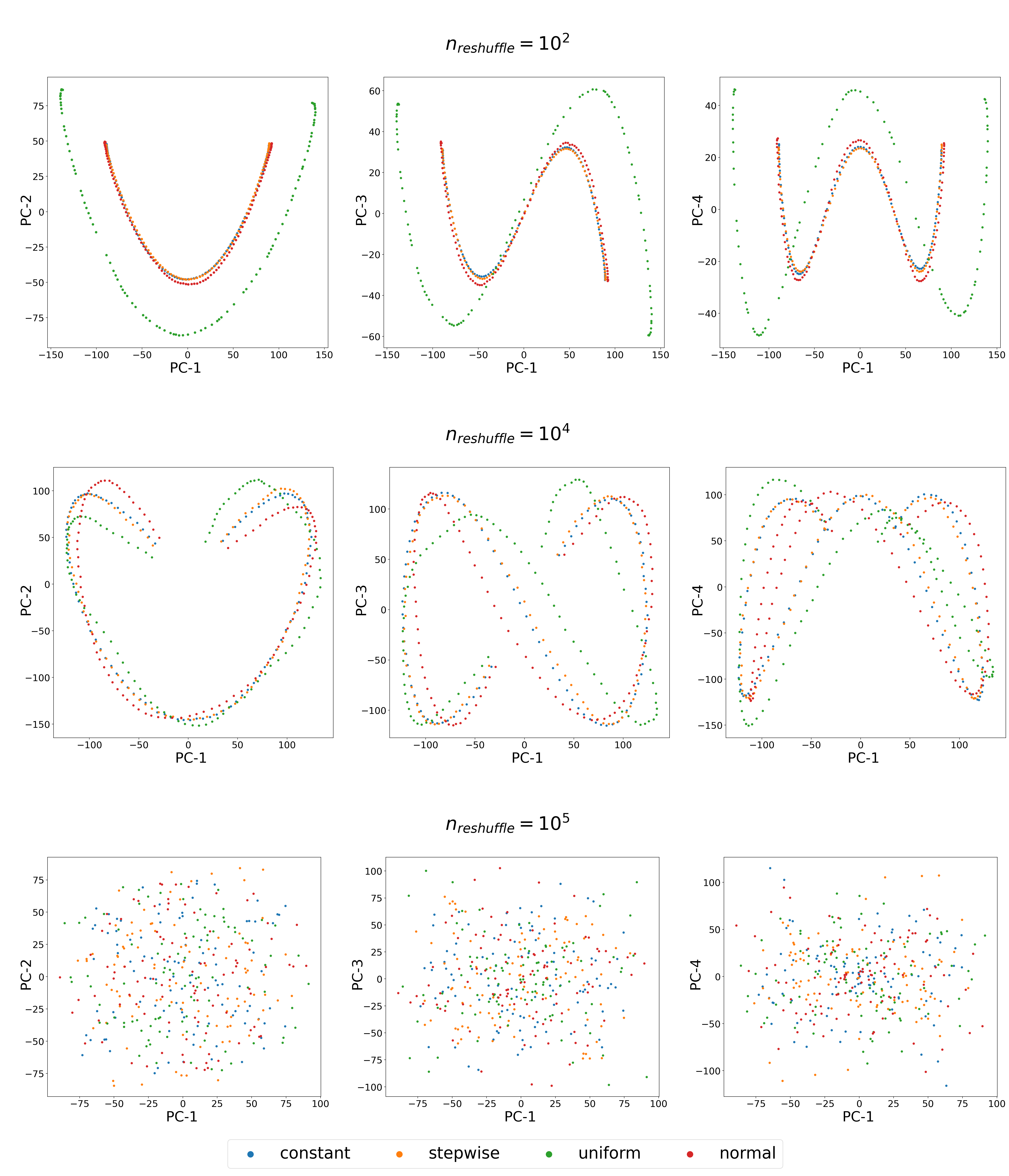}
	\caption{\label{fig:PCA-all-dist-diff-resh} Principal component analysis sets of samples obtained using the Metropolis algorithm  on $V_{{\rm (diff)},\,i}$ distributions of sections~\ref{sec:vacua-distr-constant}-\ref{sec:vacua-distr-normal}  with different values of $n_{{\rm reshuffle}}$ and all with fixed $N = 10^5$. The target value is $\mu=0.1$ in the notation of~\eqref{eq:mu-description-Vtilde}, with $\delta\tilde{V} = 10^{-10}$ in the continuous cases, and the PCA is done independently on each set.}
\end{figure}

\begin{figure}[H]
	\includegraphics[scale=.21]{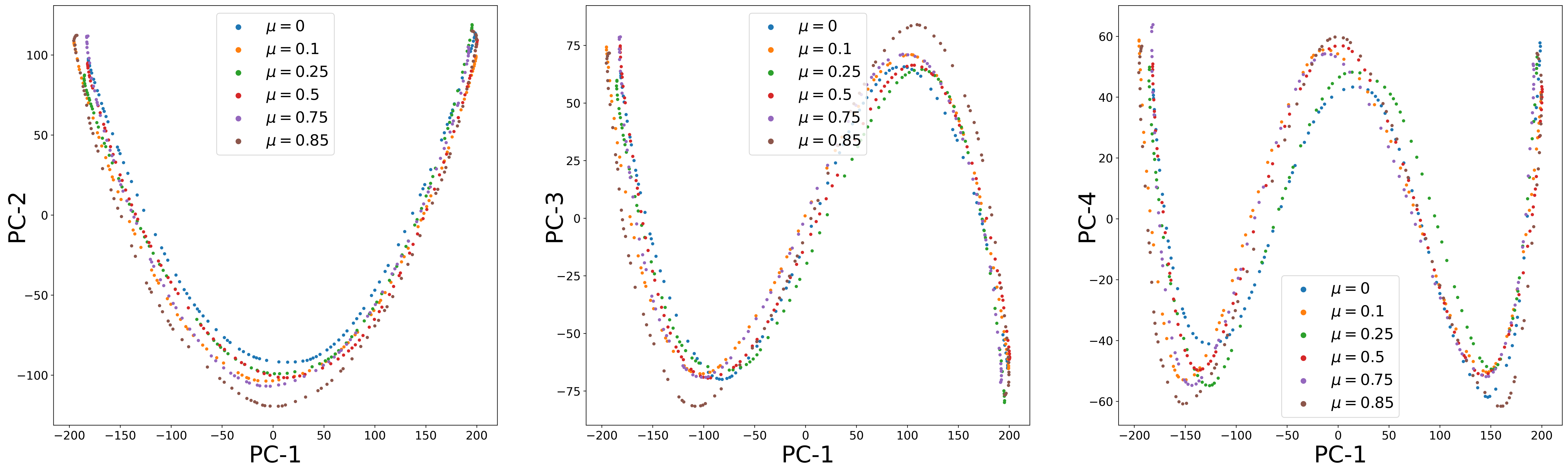}
	\caption{\label{fig:PCA-uniform-N=10^6-different-mu} Principal component analysis of a set of samples obtained using the Metropolis algorithm on the uniform distribution of section~\ref{sec:vacua-distr-uniform} with $N=10^6$, $\delta\tilde{V}_{{\rm target}} = 10^{-10}$, $n_{{\rm reshuffle}} = 100$ and different targets $\mu$ (in the notation of~\eqref{eq:mu-description-Vtilde}). }
\end{figure}

In this appendix we provide some additional information supporting the discussion in the main text. In Figure~\ref{fig:PCA-uniform-N=10^6-n=100} we show up to 10 principal components for the uniform distribution of section~\ref{sec:vacua-distr-uniform}. We observe the appearance of polynomial shapes of increasing degree. In Figure~\ref{fig:PCA-all-dist-diff-resh} we perform the same analysis as in the main text, but leaving $N$ fixed while changing $n_{{\rm reshuffle}}$. The result is equivalent to the situation where $n_{{\rm reshuffle}}$ is fixed and $N$ varies. Finally, in Figure~\ref{fig:PCA-uniform-N=10^6-different-mu} we show that the same structures appear independently of the target, given that the solutions are not sparse. 

\bibliographystyle{inspire}
\bibliography{references}

\providecommand{\href}[2]{#2}\begingroup\raggedright\begin{thebibliography}{10}

\bibitem{Bousso:2000xa}
R.~Bousso and J.~Polchinski, ``{Quantization of four form fluxes and dynamical
  neutralization of the cosmological constant},''
  \href{http://dx.doi.org/10.1088/1126-6708/2000/06/006}{JHEP {\bfseries 06}
  (2000) 006},
  \href{http://arxiv.org/abs/hep-th/0004134}{[arXiv:hep-th/0004134]}.

\bibitem{Susskind:2003kw}
L.~Susskind, ``{The Anthropic landscape of string theory},''
  \href{http://arxiv.org/abs/hep-th/0302219}{[arXiv:hep-th/0302219]}.

\bibitem{Bousso:2007er}
R.~Bousso and I.-S. Yang, ``{Landscape Predictions from Cosmological Vacuum
  Selection},'' \href{http://dx.doi.org/10.1103/PhysRevD.75.123520}{Phys. Rev.
  D {\bfseries 75} (2007) 123520},
  \href{http://arxiv.org/abs/hep-th/0703206}{[arXiv:hep-th/0703206]}.

\bibitem{Carifio:2017nyb}
J.~Carifio, W.~J. Cunningham, J.~Halverson, D.~Krioukov, C.~Long, and B.~D.
  Nelson, ``{Vacuum Selection from Cosmology on Networks of String
  Geometries},'' \href{http://dx.doi.org/10.1103/PhysRevLett.121.101602}{Phys.
  Rev. Lett. {\bfseries 121} no.~10, (2018) 101602},
  \href{http://arxiv.org/abs/1711.06685}{[arXiv:1711.06685 [hep-th]]}.

\bibitem{Douglas:2003um}
M.~R. Douglas, ``{The Statistics of string / M theory vacua},''
  \href{http://dx.doi.org/10.1088/1126-6708/2003/05/046}{JHEP {\bfseries 05}
  (2003) 046},
  \href{http://arxiv.org/abs/hep-th/0303194}{[arXiv:hep-th/0303194]}.

\bibitem{Ashok:2003gk}
S.~Ashok and M.~R. Douglas, ``{Counting flux vacua},''
  \href{http://dx.doi.org/10.1088/1126-6708/2004/01/060}{JHEP {\bfseries 01}
  (2004) 060},
  \href{http://arxiv.org/abs/hep-th/0307049}{[arXiv:hep-th/0307049]}.

\bibitem{Taylor:2015xtz}
W.~Taylor and Y.-N. Wang, ``{The F-theory geometry with most flux vacua},''
  \href{http://dx.doi.org/10.1007/JHEP12(2015)164}{JHEP {\bfseries 12} (2015)
  164}, \href{http://arxiv.org/abs/1511.03209}{[arXiv:1511.03209 [hep-th]]}.

\bibitem{Arkani-Hamed:2005zuc}
N.~Arkani-Hamed, S.~Dimopoulos, and S.~Kachru, ``{Predictive landscapes and new
  physics at a TeV},''
  \href{http://arxiv.org/abs/hep-th/0501082}{[arXiv:hep-th/0501082]}.

\bibitem{Marsh:2011aa}
D.~Marsh, L.~McAllister, and T.~Wrase, ``{The Wasteland of Random
  Supergravities},'' \href{http://dx.doi.org/10.1007/JHEP03(2012)102}{JHEP
  {\bfseries 03} (2012) 102},
  \href{http://arxiv.org/abs/1112.3034}{[arXiv:1112.3034 [hep-th]]}.

\bibitem{Bachlechner:2012at}
T.~C. Bachlechner, D.~Marsh, L.~McAllister, and T.~Wrase, ``{Supersymmetric
  Vacua in Random Supergravity},''
  \href{http://dx.doi.org/10.1007/JHEP01(2013)136}{JHEP {\bfseries 01} (2013)
  136}, \href{http://arxiv.org/abs/1207.2763}{[arXiv:1207.2763 [hep-th]]}.

\bibitem{Bao:2017thx}
N.~Bao, R.~Bousso, S.~Jordan, and B.~Lackey, ``{Fast optimization algorithms
  and the cosmological constant},''
  \href{http://dx.doi.org/10.1103/PhysRevD.96.103512}{Phys. Rev. D {\bfseries
  96} no.~10, (2017) 103512},
  \href{http://arxiv.org/abs/1706.08503}{[arXiv:1706.08503 [hep-th]]}.

\bibitem{Denef:2006ad}
F.~Denef and M.~R. Douglas, ``{Computational complexity of the landscape.
  I.},'' \href{http://dx.doi.org/10.1016/j.aop.2006.07.013}{Annals Phys.
  {\bfseries 322} (2007) 1096--1142},
  \href{http://arxiv.org/abs/hep-th/0602072}{[arXiv:hep-th/0602072]}.

\bibitem{Cole:2019enn}
A.~Cole, A.~Schachner, and G.~Shiu, ``{Searching the Landscape of Flux Vacua
  with Genetic Algorithms},''
  \href{http://dx.doi.org/10.1007/JHEP11(2019)045}{JHEP {\bfseries 11} (2019)
  045}, \href{http://arxiv.org/abs/1907.10072}{[arXiv:1907.10072 [hep-th]]}.

\bibitem{Bena:2021wyr}
I.~Bena, J.~Bl\r{a}b\"ack, M.~Gra\~na, and S.~L\"ust, ``{Algorithmically
  Solving the Tadpole Problem},''
  \href{http://dx.doi.org/10.1007/s00006-021-01189-6}{Adv. Appl. Clifford
  Algebras {\bfseries 32} no.~1, (2022) 7},
  \href{http://arxiv.org/abs/2103.03250}{[arXiv:2103.03250 [hep-th]]}.

\bibitem{Krippendorf:2021uxu}
S.~Krippendorf, R.~Kroepsch, and M.~Syvaeri, ``{Revealing systematics in
  phenomenologically viable flux vacua with reinforcement learning},''
  \href{http://arxiv.org/abs/2107.04039}{[arXiv:2107.04039 [hep-th]]}.

\bibitem{Cole:2021nnt}
A.~Cole, S.~Krippendorf, A.~Schachner, and G.~Shiu, ``{Probing the Structure of
  String Theory Vacua with Genetic Algorithms and Reinforcement Learning},'' in
  {\em {35th Conference on Neural Information Processing Systems}}.
\newblock 11, 2021.
\newblock \href{http://arxiv.org/abs/2111.11466}{[arXiv:2111.11466 [hep-th]]}.

\bibitem{Abel:2014xta}
S.~Abel and J.~Rizos, ``{Genetic Algorithms and the Search for Viable String
  Vacua},'' \href{http://dx.doi.org/10.1007/JHEP08(2014)010}{JHEP {\bfseries
  08} (2014) 010}, \href{http://arxiv.org/abs/1404.7359}{[arXiv:1404.7359
  [hep-th]]}.

\bibitem{Halverson:2019tkf}
J.~Halverson, B.~Nelson, and F.~Ruehle, ``{Branes with Brains: Exploring String
  Vacua with Deep Reinforcement Learning},''
  \href{http://dx.doi.org/10.1007/JHEP06(2019)003}{JHEP {\bfseries 06} (2019)
  003}, \href{http://arxiv.org/abs/1903.11616}{[arXiv:1903.11616 [hep-th]]}.

\bibitem{Larfors:2020ugo}
M.~Larfors and R.~Schneider, ``{Explore and Exploit with Heterotic Line Bundle
  Models},'' \href{http://dx.doi.org/10.1002/prop.202000034}{Fortsch. Phys.
  {\bfseries 68} no.~5, (2020) 2000034},
  \href{http://arxiv.org/abs/2003.04817}{[arXiv:2003.04817 [hep-th]]}.

\bibitem{Constantin:2021for}
A.~Constantin, T.~R. Harvey, and A.~Lukas, ``{Heterotic String Model Building
  with Monad Bundles and Reinforcement Learning},''
  \href{http://arxiv.org/abs/2108.07316}{[arXiv:2108.07316 [hep-th]]}.

\bibitem{Abel:2021rrj}
S.~Abel, A.~Constantin, T.~R. Harvey, and A.~Lukas, ``{Evolving Heterotic Gauge
  Backgrounds: Genetic Algorithms versus Reinforcement Learning},''
  \href{http://dx.doi.org/10.1002/prop.202200034}{Fortsch. Phys. {\bfseries 70}
  no.~5, (2022) 2200034},
  \href{http://arxiv.org/abs/2110.14029}{[arXiv:2110.14029 [hep-th]]}.

\bibitem{Loges:2021hvn}
G.~J. Loges and G.~Shiu, ``{Breeding Realistic D-Brane Models},''
  \href{http://dx.doi.org/10.1002/prop.202200038}{Fortsch. Phys. {\bfseries 70}
  no.~5, (2022) 2200038},
  \href{http://arxiv.org/abs/2112.08391}{[arXiv:2112.08391 [hep-th]]}.

\bibitem{Grana:2005jc}
M.~Grana, ``{Flux compactifications in string theory: A Comprehensive
  review},'' \href{http://dx.doi.org/10.1016/j.physrep.2005.10.008}{Phys. Rept.
  {\bfseries 423} (2006) 91--158},
  \href{http://arxiv.org/abs/hep-th/0509003}{[arXiv:hep-th/0509003]}.

\bibitem{Douglas:2006es}
M.~R. Douglas and S.~Kachru, ``{Flux compactification},''
  \href{http://dx.doi.org/10.1103/RevModPhys.79.733}{Rev. Mod. Phys. {\bfseries
  79} (2007) 733--796},
  \href{http://arxiv.org/abs/hep-th/0610102}{[arXiv:hep-th/0610102]}.

\bibitem{Martinez-Pedrera:2012teo}
D.~Martinez-Pedrera, D.~Mehta, M.~Rummel, and A.~Westphal, ``{Finding all flux
  vacua in an explicit example},''
  \href{http://dx.doi.org/10.1007/JHEP06(2013)110}{JHEP {\bfseries 06} (2013)
  110}, \href{http://arxiv.org/abs/1212.4530}{[arXiv:1212.4530 [hep-th]]}.

\bibitem{Cicoli:2013cha}
M.~Cicoli, D.~Klevers, S.~Krippendorf, C.~Mayrhofer, F.~Quevedo, and
  R.~Valandro, ``{Explicit de Sitter Flux Vacua for Global String Models with
  Chiral Matter},'' \href{http://dx.doi.org/10.1007/JHEP05(2014)001}{JHEP
  {\bfseries 05} (2014) 001},
  \href{http://arxiv.org/abs/1312.0014}{[arXiv:1312.0014 [hep-th]]}.

\bibitem{Demirtas:2019sip}
M.~Demirtas, M.~Kim, L.~Mcallister, and J.~Moritz, ``{Vacua with Small Flux
  Superpotential},''
  \href{http://dx.doi.org/10.1103/PhysRevLett.124.211603}{Phys. Rev. Lett.
  {\bfseries 124} no.~21, (2020) 211603},
  \href{http://arxiv.org/abs/1912.10047}{[arXiv:1912.10047 [hep-th]]}.

\end{thebibliography}\endgroup
\end{document}